# Probing Localized Surface Plasmons of Trisoctahedral Gold Nanocrystals for Surface Enhanced Raman Scattering

**Achyut Maity**, **Arpan Maiti**, **Biswarup Satpati**, **Avinash Patsha**, **Sandip Dhara**, and **Tapas Kumar Chini**

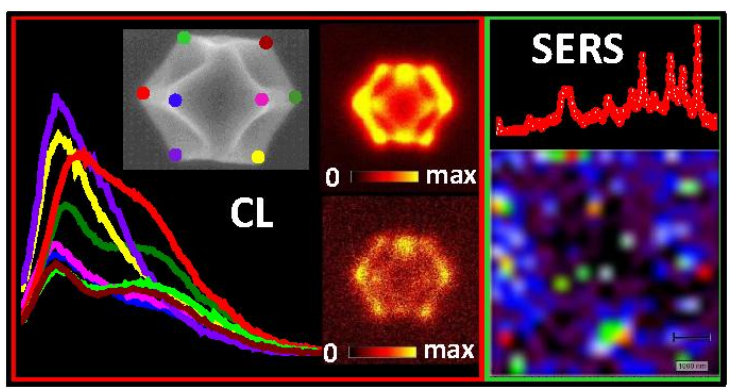

**ABSTRACT.**

Trisoctahedral (TOH) shaped gold (Au) nanocrystals (NCs) have emerged as a new class of metal nanoparticles (MNPs) due to its' superior catalytic and surface enhanced Raman scattering (SERS) activities caused by the presence of high density of atomic steps and dangling bonds on their high-index facets. We examine the radiative localized surface plasmon resonance (LSPR) modes of an isolated single TOH Au NC using cathodoluminescence (CL), with high resolution spatial information of the local density of optical states (LDOS) across the visible spectral range. Further, we show pronounced enhancement in the Raman scattering by performing Raman spectroscopic measurements on Rhodamine 6G (R6G) covered TOH Au NPs aggregates on a Si substrate. We believe that the hot spots between two adjacent MNP surfaces ('nanogaps') can be significantly stronger than single particle LSPRs. Such 'nanogap' hotspots may have crucial role on the substantial SERS enhancement observed in this report. Consequently, the present study indicates that MNPs aggregates are highly desirable than individual plasmonic nanoparticles for possible applications in SERS based biosensing.

**INTRODUCTION**.

In the search for new optical materials contributing towards localized surface plasmon resonance (LSPR), the study of metal nanoparticles (MNPs) of various shape and sizes have been a major focus of research in the area of photonics. LSPR is a result of collective oscillations of conduction band electrons in MNP and exhibits strong enhancement of near electromagnetic (EM) field intensity at the surface of MNPs, as well as, the enhancement of the far-field scattering by the particle.[1,2] The EM field enhancement from such plasmonic metal nanostructures shows remarkable applications in biosensing and bioimaging,[3–5] photovoltaics,[6,7] optical trapping[8] and surface enhanced Raman scattering (SERS),[9–13] which develop the basis of a rapidly developing field of research in nano-photonics called plasmonics.[1,14–17] In particular, SERS technique has received an intense attention in sensing, and imaging the biological, as well as, chemical species due to its capability of highly sensitive structural detection of very low-concentrated analytes.[18–22] The powerful vibrational spectroscopy technique is mainly governed by the amplification of EM fields generated through the excitation of LSP of the SERS substrate. The SERS activity crucially depends on LSP properties of MNPs on selected substrate. The noble MNPs (Ag and Au) are most commonly used for SERS activity, in the form of colloidal solution, as well as, solid substrates because of their unique optical properties. The three major factors involved in the Raman enhancement by MNPs are the size, shape and relative position of the NPs. Moreover, the particles having sharp surface curvatures[11,13,23] and tips[24–26] may provide multiple hotspots at their surface curvatures, sharp tips and also at the junction between two or more NPs. In this context, the particles having sharp surface curvatures and tips like nanoprism,[27–29] nanorod,[30] nanodecahedra,[31–33] and nanostar/flower[34,35] are exploited extensively owing to the strong localization of the EM fields at their sharp corners or tips. Therefore, it is highly demanding to understand the LSP properties of individual MNPs of different size and shape while using them for SERS activity, in the form ensemble MNPs on solid substrates.

The LSP properties of single MNP with simple geometry like rod,[30] triangles[31–33] having low-index {111} and {100} facets are well reported. Recently, a new class of Au MNP structure, namely, trisoctahedral (TOH), which is a polyhedron bounded by 24 exposed high-index facets with Miller indices of {*hkl*} *(h > l)* has drawn much attention[36–41] because of its superior catalytic,[11,42,43] and SERS activities[11,13] caused by the presence of high density of atomic steps and dangling bonds on the high-index facets. A well developed concave TOH has multitipped[13] appearance with sharp edges and corners those are favorable for possessing possible multiple hot spots. For TOH particle where the different edges, corners and facets were oriented randomly with respect to the substrate, the possibility of the existence of out-of-plane modes along with the role of substrate on the LSP enhanced photon emissions were missing in the published reports.[11,13] Consequently, experimental investigation on the spatial distribution of plasmons on single particle level is of utmost importance to understand the basic physics governing the local EM field enhancement.[29] In recent years, the use of tightly focused electron beam from electron microscope is reported to probe SPs either through detection of energy loss by the in-elastically scattered electrons transmitted through the sample, called electron energy loss spectroscopy (EELS)[16,17,31,44] in case of scanning transmission electron microscopy (STEM) or through detection of electron beam induced luminescence from the sample, called cathodoluminescence (CL)[28,29,45] in case of scanning electron microscopy (SEM). These local electron probes show excellent spatial resolution from about 20 nm (for SEM) to near atomic resolution level (for STEM),[29,31] thereby providing a unique resource for studying single particle spectroscopy. The EELS measures a quantity close to the full EM local density of states (EMLDOS). On the other hand, the CL measures a different quantity called radiative EMLDOS.[46] As the electron beam can be raster scanned in SEM / STEM, one can also obtain the information about the spatial variations of the EMLDOS. Additionally, the EMLDOS is also accessed by recording a single spectrum at a fixed point.

Here, we report for the first time the photon emission spectroscopy and imaging of individual TOH Au NC using CL technique in a SEM. The experimentally obtained spectral and photon imaging data were extensively analyzed using FDTD numerical simulation to generate electric near-field intensity maps along with the associated vectorial plots. Such analysis helped us to identify the pattern of charge oscillations along the in-plane and out-of-plane directions corresponding to different LSP modes in addition to the effect of substrate on the plasmonic properties of individual TOH Au particle. Exploiting the strong EM field enhancement by LSPR of TOH Au NPs, we demonstrate the significant enhancement SERS active substrates.

**METHODS**

**Material Synthesis.**

A seed mediated growth method was used to synthesize the TOH Au particles.[12,13] Chemicals including gold (III) chloride trihydrate ($HAuCl_4 \cdot 3H_2O$) ≥ 99.9%, sodium borohydride ($NaBH_4$ ≥ 99%) granular, 98%, L-ascorbic acid ($C_6H_8O_6$) ACS reagent ≥ 99%, were obtained from Sigma-Aldrich. (1Hexadecyl) trimethylammonium chloride (CTAC, 96%) was obtained from Alfa Aesar. The Milli-Q water was used throughout the synthesis works. At first, an ice cold freshly prepared solution of 0.30 mL, 10 mM $NaBH_4$ was quickly injected into a solution composed of 10.00 mL, 0.10 M CTAC, and 0.25 mL, 10 mM $HAuCl_4$ under magnetic stirring (1200 rpm) for colloidal Au seeds preparation. Next, this as-prepared seed solution was diluted 1000 folds with CTAC (0.10 M) solution which was further used for the seed mediated growth technique. Then, the growth solution was prepared by sequentially adding $HAuCl_4$ (0.50 mL, 10 mM) and L-ascorbic acid (1.0 mL, 0.10 M) into a CTAC (10.00 mL, 0.10 M) solution. Finally, 0.01 mL of the diluted Au seed solution was added to the growth solution and the solution was left undisturbed for overnight to complete the growth process of the TOH Au particles. As-synthesized TOH shaped Au particles were drop-casted on a mirror-polished side of a piece of cleaned crystalline Si substrate for morphological analysis using a FESEM (Zeiss, SUPRA 40). A carbon coated 3 mm grid was dipped within the growth solution and was dried for sufficient long time before inserting it in the analysis chamber of TEM for structure investigation using a high resolution transmission electron microscope (HRTEM; FEI, Tecnai G2 F30, S-Twin) operating at 300 kV.

**Cathodoluminescence Measurements.**

CL spectroscopy and imaging were carried out using Gatan MonoCL3 optical detection system attached with the ZEISS SUPRA 40 SEM.[47] The electron beam of 30 keV energy, with a current of 11.6 nA, from a hot Schottky field-emission gun of the SEM was focused onto the sample through a 1 mm opening of a paraboloid mirror with ~ 5 nm spot size. The emitted light from the sample under electron beam impact was collected by the paraboloid mirror. To ensure the maximum light collection, the sample was placed in the mirrors focal plane, which was laid approximately 1 mm below the bottom plane of the mirror. The mirror was capable to collect the emitted photon signals from the sample within a large opening angle of about $1.42\pi$ sr of the full $2\pi$ sr of upper hemisphere. The collected light signals from the sample were collimated to a 300 m Czerny-Turner type optical monochromator (with a spectral band-pass of approximately ~ 11 nm) through a hollow aluminum tube. Finally, the signal from the monochromator was directed to a Peltier-cooled high sensitivity photomultiplier tube (HSPMT). This SEM based CL system could be operated in two modes, namely, monochromatic or MonoCL mode and panchromatic or PanCL mode. The MonoCL mode was used to acquire the CL spectra serially in the wavelength range 500-900 nm with a step size of 4 nm. The dwell time was kept at 0.25s during the entire experiment. All the spectra are presented after averaging over five or six spectra for each electron beam position and then were corrected from the substrate background. The monochromatic photon maps or CL images were recorded by scanning the electron beam over the sample at a selected resonance wavelength of the CL spectrum. The monochromatic CL images were recorded with 256 × 256 pixels and the total exposure time in capturing one frame was 13.10s. In PanCL mode the emitted light signals skipped the monochromator and directly reach to the HSPMT. So, wavelength specific information was not available in the panchromatic image. The bright spot in the photon maps including monochromatic and panchromatic maps indicates the resonant excitation of the LSP modes.

**Surface Enhanced Raman Scattering Measurements**

The SERS substrates were prepared by spin coating the aqueous solution of TOH Au particles on Si substrates. Prior to the spin coating, Si substrates were treated with piranha solution to modify them as hydrophilic which helps in the uniform distribution of aqueous TOH shaped Au particles. The substrates were then ultrasonically cleaned with acetone, alcohol and subsequently dried with $N_2$ gas. TOH shaped Au coated Si substrates were then exposed to Ar-plasma for 2 min at $10^{-1}$ mbar, to burn the unnecessary carbon. Aqueous solution of Rhodamine 6G (R6G) molecules with different concentrations ($10^{-4}$M - $10^{-7}$M) were used as analyte to study the SERS activity of the TOH Au particles. Raman scattering studies were performed using micro-Raman spectrometer (inVia, Renishaw) mounted in the back scattering configuration. The samples were excited with Ar+ laser of wavelength 514.5 nm and the scattered signals were collected using 100X objective. A grating of 1800 lines/mm and thermoelectrically cooled CCD detector were used to analyse and record the spectra.[48] The spectra were acquired at different places on the sample to make sure the homogeneous and reproducible SERS activity by the TOH Au particles. The SERS imaging was performed over a pre-defined area and grid resolution using fully automated motorized sample stage (Renishaw MS20), having a spatial resolution of 100 nm.[49] In the present study, a total area of 6 μm × 6 μm with 300 nm grid resolution is probed for the SERS mapping.

**FDTD Simulations**

3D-FDTD simulations (from Lumerical solutions, Canada) were carried out to understand the electron beam induced LSP responses including the charge distribution, near-field intensity maps and substrate's effect of the TOH NP. In FDTD, the Maxwell's equations are solved in discretized time and discretized space to follow the response of a material to any external electromagnetic (EM) field (like the associative evanescent wave with electron beam in case of CL). In FDTD simulations, the electron beam is modelled with a large number of closely spaced dipoles placed vertically with a temporal phase lag that is associated with the electron velocity. The line current density of the electron beam can be written as,

$$\vec{J}(\vec{r},t) = -eV\hat{u}_Z\delta(Z - Vt)\delta(X - X_0)\delta(Y - Y_0) \dots \dots \dots \dots \dots \dots \dots \dots \dots \dots \dots (1)$$

where $e$ is the electronic charge and $V$ is the velocity of electron, ($X_0$, $Y_0$) represents the position of the electron beam and $Z$ is the direction of electron velocity and $\hat{u}_Z$ is the unit vector along $Z$ direction. The phase lag factor between two consecutive dipole is defined as ($Z/V$), here $V = 0.32c$ corresponding to the 30 keV electron energy used in the present present CL experiment, with $c$ being the velocity of light in free space. In the absence of any structure, electron beam moving at a constant velocity is unable to generate any radiation. However, in FDTD, we are obliged to simulate only a finite portion of the electron beam path, and the sudden appearance and disappearance of the dipole beam will induce radiation. To solve this issue, we operated a second reference simulation removing all the structures. Then the EM fields at angular frequency ω were calculated by taking the difference in fields between the two simulations. The geometric parameters like lengths, curvature of the corners were measured from the HRSEM images, and then were used for modelling of the Au TOH structure.

The dielectric constant/refractive index for gold in the wavelength range 500-900 nm was used from the data tabulated by CRC Handbook of Chemistry and Physics[50] which were fitted (Figure S1 in Supporting Information (SI)) using a multi-coefficient models (MCMs),[51] as the Lorentz, Drude and Debye models were often insufficient for real materials. MCMs deal with an arbitrary dispersion of a dielectric function through the more extensive set of basis functions. For FDTD simulations, the modelled TOH Au particle of edge length (Figure S2 in SI) 125 nm was assumed as a combined structure of one regular octahedron and eight triangular pyramids (considering equilateral in nature of triangular bases), where the triangular pyramids were attached on eight {111} facets of the octahedron by 'pulling out' their apexes from the center of {111} facets.[13] Here, the edge length of the modelled TOH structure was defined by the edge length of each triangular pyramid. The edge length and the length of a side of triangular base of each pyramid were assumed to be 125 nm and 190 nm, respectively. According to the geometry of TOH structure, the edge length of octahedron is equal to the length of a side of the triangular base (i.e., 190 nm) of triangular pyramid. However, we have only considered the mesoscopic structural geometry in the modelling due to the limited computational resources in terms of memory and speed. The Si substrate was assumed as a non-dispersive material (with fixed dielectric constant of 4) within the wavelength range 500-900 nm. The dimension of the Si substrate was considered to be 5 μm × 5 μm × 4 μm in the present calculations. Here in our case, the mesh override region (2 nm × 2 nm × 2 nm in $X \times Y \times Z$ direction) and FDTD mesh accuracy of 3 were kept fixed during all simulations constrained by our limited computational resources in terms of memory and speed.

In order to clarify the particle morphology and orientation of different points in presence and absence of substrate, we captured the 3D perspective view of the TOH shaped particle directly from the 3D FDTD software package layout used in the present modelling. Figure 1a shows the orientation of the TOH morphology at free standing configuration (in free space) where the substrate is absent. The Cartesian coordinate axes defined for electron beam (e-beam) are designated by *X, Y, Z* (grey color) while the co-ordinate axes attached with the particle are designated with *x, y, z* (green color). For example, the incoming electron beam is indicated with the red arrow parallel to the *Z* axis. At free-standing configuration, both the coordinate frames coincide to each other and the points like A-F belong to the XY plane. On other hand, at experimental condition, while the particle is at rest on the substrate, two apex points and one base point (making a triangular plane) must have to be in contact with the substrate (Figure 1b) making the particle coordinate system being tilted with respect to the free standing configuration. By analyzing the

SEM image of the TOH Au particle with the FDTD modelling we could determine the tilt angle of 7.5º between the *Z* axis and *z* axis while the particle is at rest on the substrate. Consequently, the points like A, D are nearer to the substrate than the points like B, C (Figure 1b). The amount of this tilt angle will depend on the height of the triangular pyramid as well as the size of the TOH Au particle.

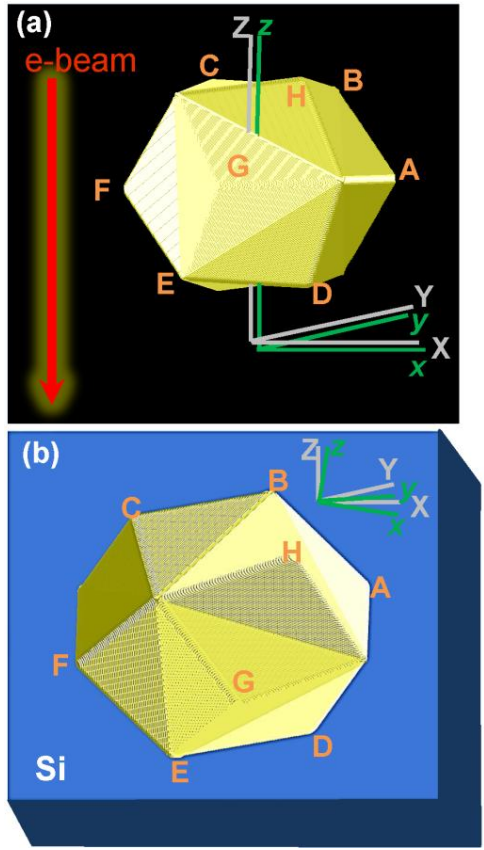

**Figure 1:** 3D perspective view of the TOH shaped particle modelled in the present FDTD software package layout in absence (a) and presence (b) of silicon (Si) substrate.

For the normal SEM operation the sample mounting stage remains horizontal and one sees the top view of the particle from the SEM image. To understand the three dimensional (3D) perspective of the particle morphology and different facets or triangular pyramidal parts one needs to tilt the sample stage. See the Figure S3 in SI for a representative SEM image at 45° tilt angle of the sample stage. However, during CL experiment one has to insert the light collecting paraboloid mirror for which the sample stage is constrained to remain at horizontal position, a condition imposed by the maximum light collecting condition (see Cathodoluminescence Measurements section). This means, during CL experiments, while the sample is at rest on the substrate, the angle between $Z$ and $z$ is always 7.5° for the present TOH Au particle.

## RESULTS AND DISCUSSION

**Cathodoluminescence Analyses**

Figure 2a shows a representative SEM image of as-synthesized Au TOH nanostructures. The NPs are mostly aggregated in clusters. However, a single isolated TOH Au NP of edge length 125 nm (encircled with the red box), far from the clusters, is selected to investigate the LSP properties of such complex shaped MNP. Geometrically, the TOH particle is a crystalline structure where eight triangular pyramids are attached on the eight {111} facets of the octahedra by pulling out their apexes from the center of {111} facets. One of the pyramidal parts is very prominent in the tilted stage configuration of SEM imaging (Figure 2b). Figure 2c shows bright field (BF) TEM image of an aggregate of TOH particles in random orientations. In Figure 2e, we show the selected area electron diffraction (SAED) pattern from an isolated TOH particle (Figure 2d). The SAED pattern of the TOH NP can be indexed to the [011] zone axis of a single crystal of fcc Au; suggesting that the TOH NPs are single crystals,[36,37] rather than multiply twinned crystals. Figure 3a shows the experimentally acquired site-specific CL spectra from different beam injected points marked as A-H.

Now, the points namely, A, B, C and H are symmetric to the points D, E, F and G respectively, along the two-fold rotational symmetry (Figure S2 in SI).[40] The spectra were collected in two different wavelength ranges, 500-700 nm and 700-900 nm, to minimize the sample drift. Finally, the spectra were merged with proper normalization.

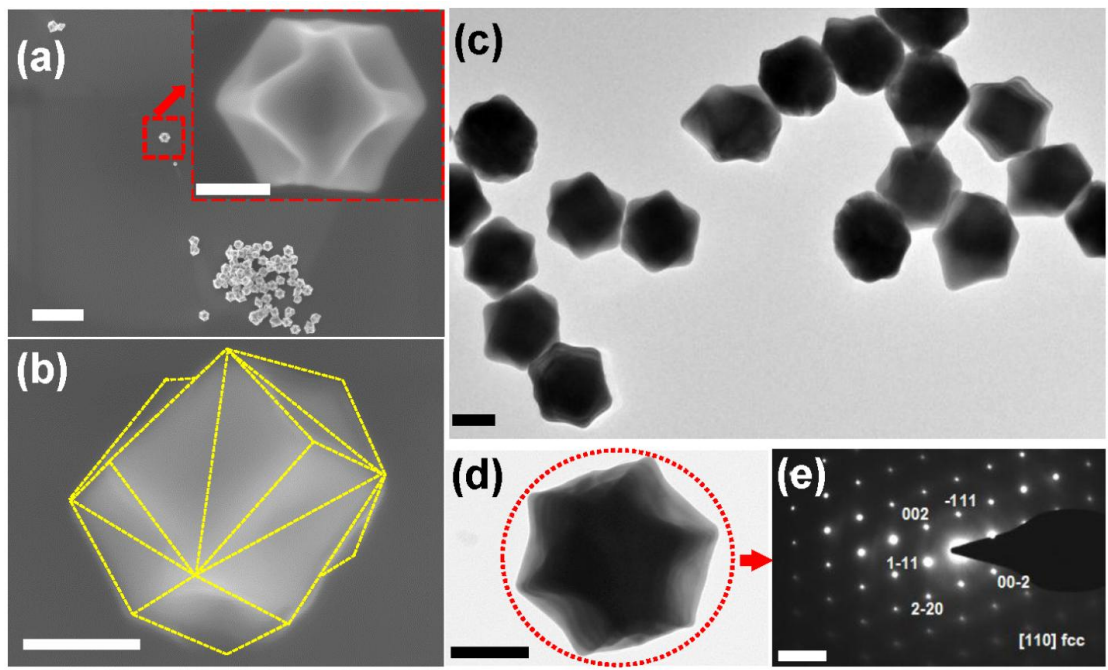

**Figure 2:** (a) SEM image showing the aggregation of TOH Au particles on Si substrate. The studied particle is indicated with red colored box. The HRSEM image of selected TOH is shown in the inset figure. The scale bars in the figure and inset figure are respectively 1 μm and 100 nm. (b) SEM image of the TOH Au particle in tilted configuration of mounting stage. The scale bar is 100 nm (c) Bright field TEM image of a collection of TOH Au particles. An isolated TOH Au particle (encircled with red circle) and its corresponding selected area electron diffraction (SAED) pattern are shown in (d) and (e) respectively. The scale bars in the figures (c), (d) and (e) are 100 nm, 100 nm and 5 nm$^{-1}$, respectively.

Finally, the spectra were merged with proper normalization. It is interesting to note that except the points A, D electron beam impact at all other selected points gives rise to double peaked emission spectra. Moreover, the presence of lower wavelength resonant peak at around 548 nm in the CL spectra is common for electron beam excitation at all the marked points (A-H) showing gradual decrease in intensity as the beam probes from the symmetry points located nearer to the farthest from the substrate, specifically from A to C (see FDTD Calculations in Methods section and Figure S3 in SI), except the point E for which the lower wavelength peak arises ~~at~~ around 560 nm. Consequently, the highest intensity peaks at around 548 nm is observed for the electron beam impact on the points A and D (spectra with yellow and violet color in Figure 3a). For electron beam impact at B and E, the higher wavelength resonant peak occurs at around 650 nm. However, for beam impact at the points C and F the higher wavelength resonant peak occurs at around 670 nm. Remarkably, full-width at half maxima (FWHM) of the lower wavelength resonant peaks are less than half of that for the higher wavelength peaks, meaning lower wavelength resonant peaks are significantly narrower than the higher wavelength peaks (Table S1 in SI). The broadening in all the spectra appear due to the increased damping.[13,17]

Although we present here the experimental spectral data for a single particle, the reproducibility of our results is checked by more measurements (Figure S4 in SI). The spectral analysis of the experimentally observed CL resonant peaks corresponding to different LSP modes are corroborated with the spatial information through the wavelength specific photon maps or monochromatic CL images shown in Figures 3e-h preceded by a panchromatic CL map (Figure 3d). The plasmon induced luminescence is already discernible in the panCL image (Figure 3d) showing regions of high intensity emission at the points, marked as A-H. One can easily observe from the monochromatic photon maps at 548 and 560 nm (Figures 3e and 3f) that the luminescence features are nearly same in both the photon maps.

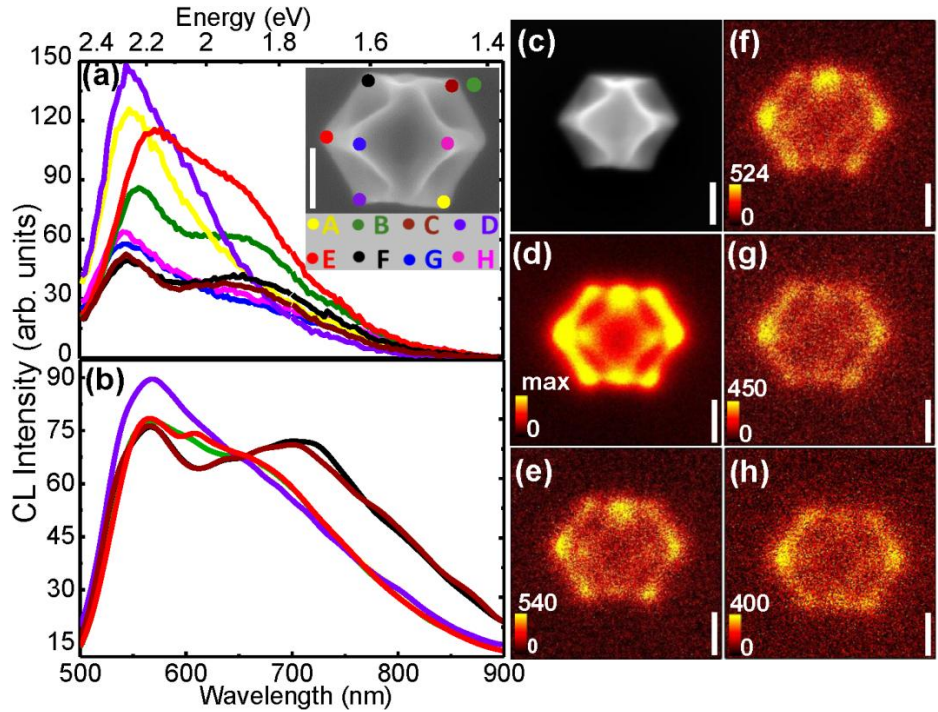

**Figure 3:** (a) CL spectra from selected points marked as A-H of the studied TOH Au particle. The inset SEM image shows the electron beam impact points with different colored dots. (b) 3D-FDTD simulated spectra of the modeled TOH structure for beam impact at the points (A, B..., F). The electron beam is kept in 1 nm away from all the selected points during all numerical calculations, unless mentioned specifically. The SEM image and the corresponding panchromatic CL map/image of the same morphology are shown in (c) and (d) respectively. (e)-(h) Monochromatic CL maps/images at wavelengths 548, 560, 650 and 670 nm, respectively. Scale bar is 100 nm in all images.

Similar observation is also made for the monochromatic CL images of Figures 3g and 3h corresponding to the peaks 650 and 670 nm. This apparent small shift between maxima may just as well be due to the overlap of two resonant line shapes with varying weights. We thus have basically two LSP modes, one corresponding to the resonant peaks 548/560 nm and other corresponding to 650/670 nm.

The FDTD calculated spectra of the modelled TOH Au particle are shown in Figure 3b for all selected points except G and H. In modelling, the TOH Au particle is placed on Si substrate at rest making an angle of 7.5° (see the FDTD Simulations in Methods section) between the $Z$ axis and $z$ axis reflecting the experimental configuration during CL measurement. Simulated analyses reveal that the lower wavelength LSPR mode always appears at 560 nm for all the selected points. Whereas, experimentally it appears either at 548 nm or 560 nm. Interestingly, while comparing the calculated spectra with the experimental one, the best match with the experiment is found for the beam impact at the points B and E (green and red curve in Figure 3b). The small additional local maxima observed in the calculated spectra arise due to numerical artifact.[51] The experimentally acquired CL spectrum from point B and E should have been same due to symmetry reason. However, the experimental results show that there is a slight mismatch of ~12 nm in the lower wavelength LPSR peak positions. The reason can be explained as follows: It is well reported in the published literature[33,52] that the position as well as the intensity of any particular LSPR mode depends on the curvature and concaveness of the particles morphology among other factors. However, they are not exactly same for both the points as is evident from the HRSEM images (Figure 2a) rendering the difference in the spectra taken at these two points. In case of numerical modelling, we maintained equal roundness of curvature of 5 nm in all corner regions of the TOH Au particle. Consequently, the calculated spectra from both these points do not show significant difference. Mismatch between the experimental and simulated spectra appears to be more pronounced for the higher wavelength mode compared to the lower wavelength mode. As for example, for the electron beam impact at points C and F (brown and black curves in Figure 3b) where the calculated higher wavelength peak (720 nm) red shifts by 50 nm w.r.t the experimental high wavelength peak (670 nm). However, the calculated lower wavelength resonant peak (560 nm) red shifts only 12 nm compared to the experimentally observed peak at 548 nm for the CL spectra originating from points A and D (yellow and violet curve in Figure 3b). The mismatch between experimental and simulated spectra can be attributed as follows: the morphology of the selected TOH Au particle (shown in inset image of Figure 2a) in CL experiment is concave in nature. This concaveness is not perfectly maintained in the modelled TOH structure which introduces the red shift nature in the simulated spectra.[13] Additionally, the uncertainties in morphological details like surface roughness, roundness of curvature, as well as, the uncertainties in the dielectric constant, assumed in model calculation with those actually prevalent on the experimental Au particle where the possibility of ultrathin layer of surfactant cannot be ruled out; may also play a role. Overall, the fairly good agreement between experimental and simulated results indicates that the simplified modelling of the TOH shaped NP is qualitatively correct.

To get clear information about the origin of LSP modes and the corresponding oscillations of induced charges, we have performed 3D-FDTD numerical simulations in free-standing configuration without the Si substrate. For the sake of clarity an origami model of a typical TOH structure marked with the selected points of electron beam impacts A, B, H is presented in Figure 4a. Due to the two fold rotational symmetry[40] of TOH Au particle, the points A, B and C are equivalent to the points D, E and F, respectively. In order to calculate the near-field electric intensity ($|E|^2$) maps and the related electric field vector plots at various resonant peaks, two mutually perpendicular near-field power monitors are placed parallel to XY and XZ planes as depicted in Figure 4b. Here, Figure 4c presents the low magnification SEM image of an aggregate of randomly oriented TOH Au NP out of which a high magnification SEM image of an isolated TOH Au particle is selected in Figures 4d,e to define locations of the base and apex points to be used in our model calculation. The base point is a common point where the bases of four nearest neighbor pyramids meet together. In our case, point B (or E) is a base point and points A, C, H (or D, F, G) are the apexes of three nearest pyramids. The fourth apex point cannot be seen, as it is situated in the lower part of the particle.

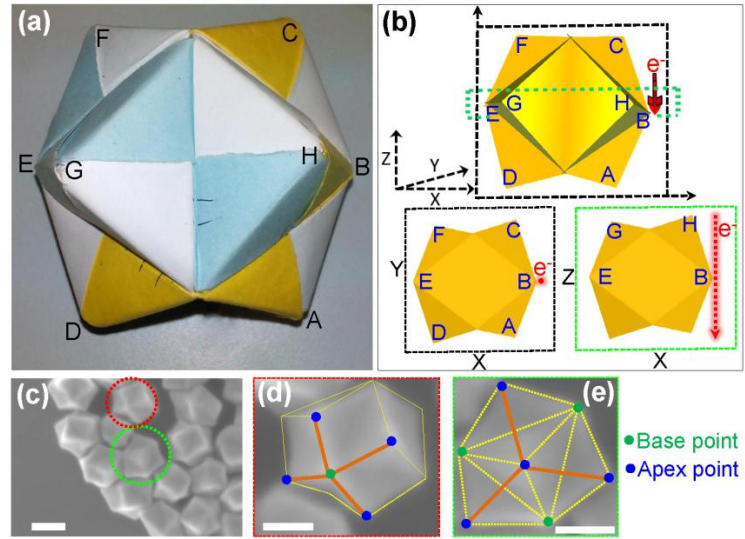

**Figure 4:** (a) Image of the origami model of a typical TOH structure. (b) Schematic diagram of the TOH Au particle in free-standing condition. Power monitors are placed at XY and XZ planes. Cross-sectional view of TOH Au particle in XY and XZ planes are shown in the lower panel of (b). Under free-standing condition, the XY plane is parallel to the equatorial plane of the particle, i.e., perpendicular to electron-beam direction and the XZ monitor is parallel to the electron beam. (c) SEM image of ensemble of TOH Au particles, where the particles are orientated in different directions. The scale bar in (c) is 100 nm. To visualize the base and the apex points, two isolated particles are selected and encircled with red and green circles. The magnified SEM images of the selected particles are shown in (d) and (e). (d) A typical schematic of base point, where the bases of 4 nearest triangular pyramids meet together. (e) A typical schematic of apex point. The scale bar in both the figures (d) and (e) is 50 nm.

We will only focus on the electron beam induced LSP responses from the points A, B and C only in free-standing configuration for which the calculated CL spectra are shown in Figure 5, indicating the existence of two LSP modes at 550 nm and 820 nm. While the experimentally observed low wavelength LSP mode (548/560 nm) agrees well with that for free-standing modelled TOH Au particle, the observed high wavelength LSP mode (650/670 nm) is blue shifted substantially with that for free-standing case, which will be discussed later in the context of substrate effect.

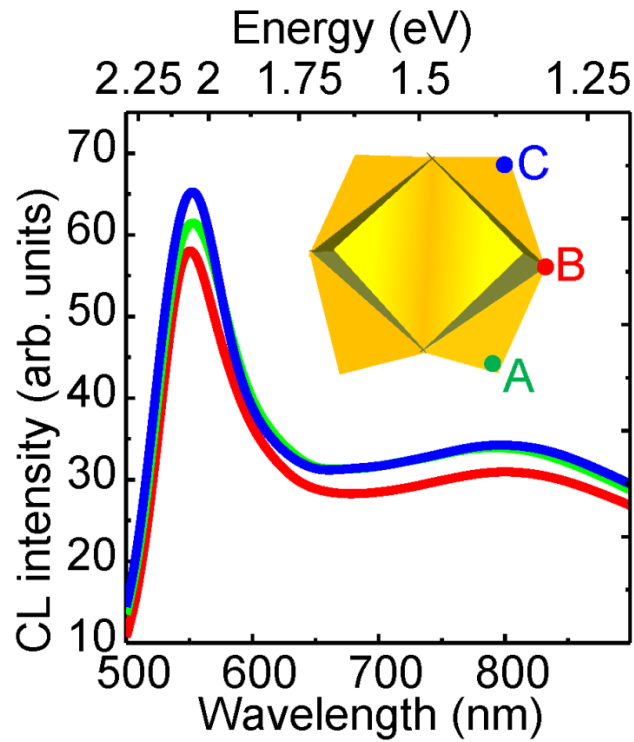

**Figure 5:** 3D-FDTD simulated CL spectrum from the selected points of the TOH structure for modelling under free-standing configuration. The spectra are calculated within the 500- 900 nm wavelength range. The beam injected points and the corresponding spectra are highlighted with same color. The schematic of the TOH model is shown in the inset.

The nature of the plasmon modes is assigned by observing the standing wave formation through the calculated near electric field intensity maps and the direction of the corresponding electric field vectors.[28–30,32,33] Electric field converging and diverging points are designated as the negatively and positively charged points, respectively.[28–30,32,33] Now, for electron beam excitation at point B, the vector plot of Figures 6a and 6b clearly depicts that the LSP mode at 550 nm has dipolar nature of oscillation both in-plane as well as out-of-plane directions. The charges of opposite polarities (±) are observed to be distributed from base point to nearest apex point, *e.g.*, between the points B-A / B-C (in XY plane) or B-H (in XZ plane). The corresponding near-field intensity map illustrates the localized and strong electric field enhancements near the points A, B, C and H. However, for electron beam excitation at point A or C, the dipolar charge oscillation between base and apex for the LSP mode at 550 nm occurs in the in-plane direction (XY plane) as revealed from the vector plots of Figures 6c and 6d. Although it seems that the field lines are emanating from the point F to C or D to A at 550 nm wavelength for electron beam excitation at point C or A (Figures 6c and 6d); however, they may also emerge from any other base points which are situated at lower portion of the particle and they are also connected with point C or A. The near-field electric intensity map also shows the strong enhancements at the negatively and positively charged regions. Interestingly, as the induced EM field from the electron beam may extend over the entire nanoparticle under resonance condition,[28] one can also observe the field enhancement at other points as well (i.e., point E in the present case) of the nanoparticle. However, the other points (i.e., the point E) do not take part in standing wave loop formation for 550 nm base-apex LSPR mode when the electron beam is probed at point B. On the other hand, for electron beam excitation either at point C or A, the higher wavelength (or lower energy) LSP mode at 820 nm also exhibits dipolar nature. However, the charge pattern oscillations occur between two nearest apex points (e.g. A-D, C-F) of TOH Au particle, as seen from Figures 6e and 6f. Geometrically, any apex of a TOH Au particle has three nearest neighbor apex points (shown in Figure 4e) owing to the three fold rotational symmetry.[40] This leads to a signature of 3-fold degenerate dipolar mode at the wavelength of $\lambda = 820$ nm in free-standing configuration for electron beam probe positioned at one of the symmetric apex points.[31,32] The distance between two nearest apexes, namely, A-D or C-F is 150 nm, as measured from inset SEM image of Figure 3a. It is greater than that between a nearest apex and a base point (B-A or B-C). Consequently, the apex-apex dipole active mode appears at higher wavelength of 820 nm compared to the dipolar LSP modes at 550 nm along the apex-base (A-B, B-C etc.) edges.

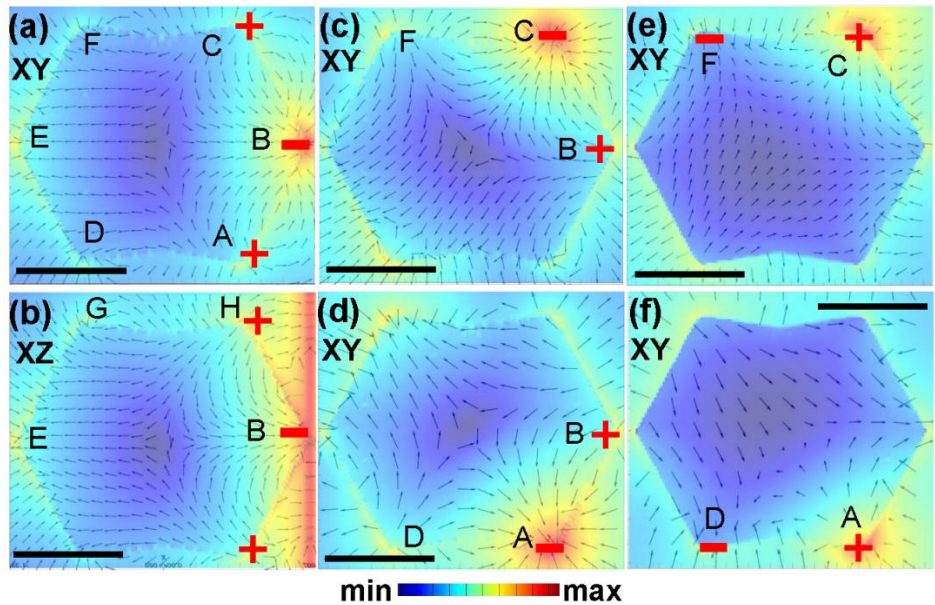

**Figure 6**: 3D-FDTD simulated near-field intensity ($|E|^2$) and corresponding field vector maps at resonant wavelengths 550 nm (a)-(d), and 820 nm (e), (f). Scale bar is 100 nm.

We now discuss the role of substrate on the observed spectra. Comparing the calculated scattering spectra of TOH Au particle for free standing condition of Figure 5 with the calculated spectra of Figure 3b for particle configured with the experimental situation, one can get a hint on the role of substrate, specifically, for the lower energy degenerate apex to apex dipole active mode. In general, the longitudinal in-plane dipolar component of particles with simple geometry like rod, triangular prism etc.[28–30] shows the redshift with respect to the free space resonance wavelength in presence of high index substrate. However, for TOH Au particle, the dipolar LSP modes are not so well defined in longitudinal and transverse directions as in the case for a particle with simple rod like morphology. The LSP modes are oriented along different directions due to the 3D orientations of different apex points. The modal distribution analyzed for e-beam injections at point B and C in the present case is much more complex than a simple elongated particle. Additionally, the losses due to the interband electronic transitions in Au, may play an important role in LSPR responses,[53] especially in the visible and ultra-violet (UV) spectral ranges (mainly 500-600 nm) when the Au TOH particle is exposed to the highly energetic electrons. In the present work, we have considered the dielectric permittivity tabulated by CRC[50] for the modeled Au TOH nanoparticle during the simulations where the losses are not taken into account. Moreover, the simulated results could be suffered due to the imperfect experimental structure or the meshing in FDTD simulations along with the transition losses.[54] Although the FDTD simulated results provide a qualitative perception about the substrate mediated interaction[31,32,44,55,56] on apex-apex dipolar LSP mode, quantitative assessment of the blue shifted nature of the apex-apex mode is difficult due to intricate morphology of the complex shaped TOH Au particle on substrate. However, the apex-apex mode is not present in both experimental and calculated spectra (Figures 3a and 3b) for electron beam excitation at points A and D. This is because, the apex points, A and D, are very close to the high index substrate like Si compared to any other points and the apex-apex LSPR mode corresponding to these points will tend to radiate mostly into the substrate.[32] It is also observed that the base-apex LSP modes like A-B, B-C do not show any noticeable shift in spectra when the substrate is taken into account. This is consistent with the monochromatic CL map corresponding to the base-apex mode either at 548 nm or 560 nm showing (Figures 3e or 3f) that the strong luminescence coming from the base and apex points of the TOH Au particle. Recently, in a few publications,[13,42] it is reported that the quadrupolar nature of the lower wavelength LSPR mode of Au TOH NPs may appear along with the dipolar one for the increased size of the TOH Au particle. It is established for a single particle that the higher order modes (like quadrupolar) are less affected by the substrate compared to the dipolar modes,[28] which is consistent with the present analyses where the substrate effect is negligible for the lower wavelength LSP mode with respect to the free standing case. Insignificant substrate effect is also reported[33] even in case of a mixed mode, which appears at higher energy or lower wavelength region of spectrum. Moreover, the broadening of a LSPR peak is generally regarded as the mixing of different plasmon modes for complex shaped nano-objects studied by CL spectroscopy.[31] The broadening of the 550 nm LSP mode may suggest a certain probability of the presence of quadrupolar mode in our result.

**SERS Analyses**

The strength of the LSPR of TOH Au NPs in SERS activity was studied by collecting the Raman spectra of Rhodamine 6G (R6G) molecules, dispersed on SERS substrates containing ensemble of Au NPs. Since it is difficult to probe the SERS activity of single TOH Au particle having size below the optical resolution of a microscope, the SERS studies were carried out on ensemble Au NPs. Though SERS is measured on the ensemble of NPs, the hot spot at the individual NPs or in the nanogap of two NPs will be primarily responsible for the ambiguity free measurement of the Raman enhancement. Generally SERS measurement is understood as a far-field measurement technique; however the information emerging from a localized interaction of the molecular dipoles with the near-field evanescent wave generated by the interaction of electromagnetic excitation and Au NPs remains intact even in the far-field.[57] An optical view of the SERS substrate with uniformly dispersed TOH Au NPs is shown (shown in inset, Figure 7). The aqueous solution of $10^{-6}$ M concentration of R6G molecules was dispersed on both SERS and pristine Si substrates to compare the spectral enhancement. The prominent Raman modes observed at 611, 773, 1123, 1188, 1307, 1364, 1423, 1509, 1573, and 1650 cm$^{-1}$ (Figure 7) are corresponds to R6G molecules on SERS substrate.[21,58] A broad peak around 958 cm$^{-1}$ corresponds to the second order Raman mode of Si substrate. All the Raman modes of R6G molecules on SERS substrate have shown the enhancement in intensity, as compared to that of the pristine Si substrate. The highest enhancement was observed for 1650 cm$^{-1}$ mode. Considering the effective laser excitation volume of 514.5 nm laser ($V = \pi r^2 \delta = 15.32$ μm$^3$; where $r$ being radius of laser spot, $\delta$ being depth of focus) by 100X objective (N.A = 0.85) and the surface area of a single TOH Au NP, the enhancement factor (EF) of a Raman mode was estimated[11,13] by, EF = ($I_{SERS}$ $N_{normal}$ / $I_{normal}$ $N_{SERS}$).

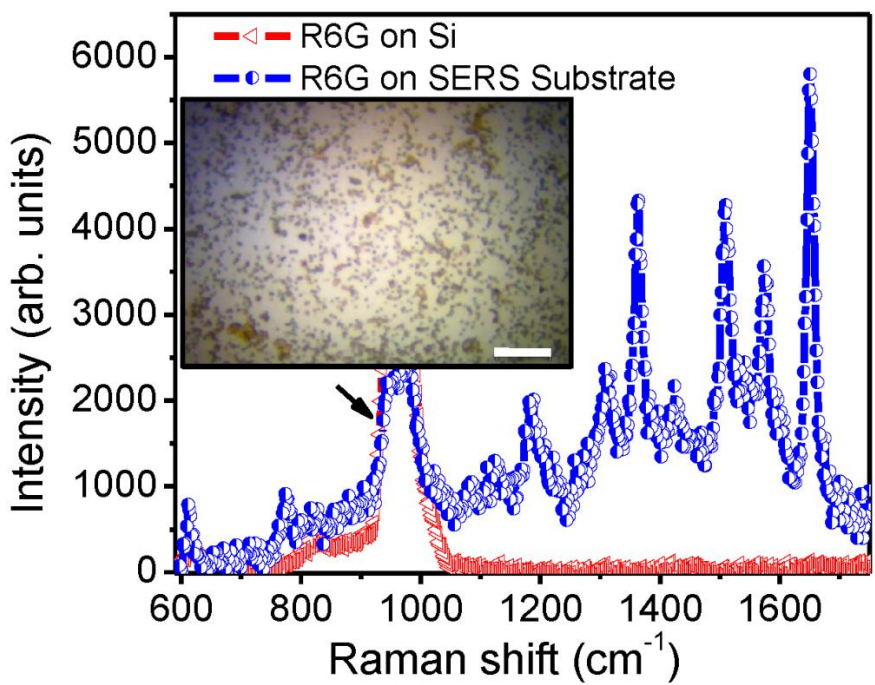

**Figure 7:** SERS spectra of R6G molecules in aqueous solution of 10$^{-6}$ M concentration dispersed on SERS substrate and pure Si substrates. Arrow mark denotes the second order

Raman mode of Si substrate. Inset represents the optical image of the SERS substrate at 100X magnification. Scale bar is 10 µm.

$I_{SERS}$ and $I_{normal}$ are the intensity of particular Raman mode observed in the SERS and normal spectra, respectively; $N_{normal}$ and $N_{SERS}$ are the number of R6G molecules in the excitation volume for the normal Raman acquisition and number of adsorbed molecules on individual TOH Au particle, respectively. $N_{normal} = (V\rho/M)N_A$ is estimated to be $2.4 \times 10^{10}$. Here $\rho$ (1.26 g.cm$^{-3}$) and $M$ (479.02 g.mol$^{-1}$) represent the density and molar mass of R6G molecules and $N_A$ being Avogadro number. Assuming monolayer formation of R6G molecules (molecular foot print size $\approx 2.2$ nm$^2$) on TOH Au particle (surface area is calculated $480 \times 10^3$ nm$^2$), we have estimated the $N_{SERS}$ to be $216 \times 10^3$. The estimated EF for the 1650 cm$^{-1}$ mode is found to be of the order of $10^8$. The observed significant enhancement in the Raman scattering on TOH Au particles may be the result of hotspots present at the tips of each particle (as shown in Figures 3d-h and Figure S5 in SI), as well as in between the particles adjacent to each other. The enhancement in the present study can be understood by invoking a general principle of energy conservation, for which the total electromagnetic energy per unit volume containing several nanoparticles should be same for any shape. In such situation, due to the strong nonlinearity of the SERS enhancement factor, one strong hotspot may lead to higher SERS enhancement than many weaker hot spots. It is known that on-resonance excitations of Au NPs generate stronger field strength needed for Raman enhancements than the off-resonance excitation. However, the Raman scattering cross-section and molecular resonance for particular molecule depend on the excitation wavelength. In this context, we have investigated the SERS of R6G by choosing the resonance excitation of TOH Au NPs, which is close to 785 nm (Figure S6 in SI). The estimated EF, however, is one order less ($10^7$) compared to that of 514.5 nm excitation.

The SERS activity of the TOH Au NPs was tested for wide range of molar concentrations ($10^{-3}$ to $10^{-8}$ M) of R6G and found that the TOH Au NPs are able to detect the R6G molecules of concentration up to $10^{-7}$ M (Figure S7 in SI). Further, the areal distribution of enhancement regions on the SERS substrate was mapped by performing the SERS imaging of R6G molecules. An area of 6 μm × 6 μm on SERS substrate containing R6G of concentration $10^{-6}$ M, was chosen for imaging (Figure 8a). The mapped integral intensity distribution of 1650 cm$^{-1}$ mode of R6G (Figure 8b) shows the high intensity spots randomly distributed over the mapped area. Interestingly, one can notice from the optical image of the Au NPs on SERS substrate as shown in Figure 8a and also from SEM image of Figures 2a, c and Figure 4c that the Au NPs are mostly in the form ensembles implying the existence of the "nanogaps" (< 100 nm) between the Au TOH NPs. Meaning, we can not unambiguously say that maximum intensity regions in the SERS map correspond to the location of hot spots where field enhancement occurred by the LSPR of single TOH Au NPs because of the diffraction limited lateral resolution of Raman microscope (~370 nm in the present case) (Figure 8b).

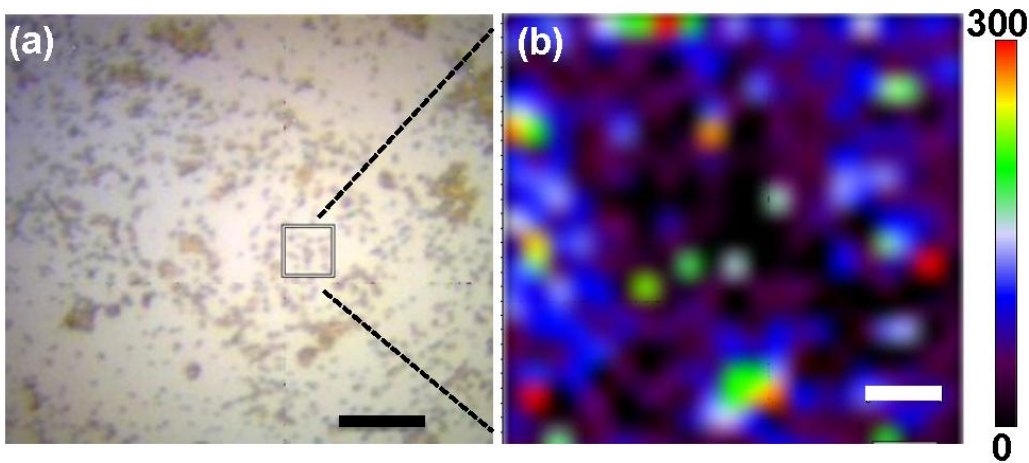

**Figure 8:** Optical view of SERS substrate containing R6G molecules of concentration $10^{-6}$M, used for SERS imaging. (a) The square box indicates the area (6 μm × 6 μm) chosen for SERS

imaging of R6G. Scale bar is 10 μm. (b) The SERS map of 1650 cm$^{-1}$ mode of R6G. Scale bar is 1 μm. The color scale bar indicates the intensity profile of the Raman mode.

However, due to the strong plasmonic coupling effects,[59] the nanogaps between the adjacent particles of the ensembles on substrate may serve as hot spots with gigantic local-field enhancements of several orders of magnitude higher than those achievable for LSPR of individual NP. We believe that such a strong hot spot could be responsible for the significant SERS enhancement observed for present TOH particles. Apart from it, the modal distributions as well as the mixing or interference of the particle plasmon modes[25] in case of anisotropic and complex shaped finite nanostructures also play very crucial role in optical properties including SERS activity of the particles. The higher order plasmon mode, i.e., the quadrupolar mode which is supposed to be present to some extent as described in our CL analysis is generally of nonradiative nature. But they can absorb light and decay through evanescent near-field, and thus may also contribute to SERS enhancements.[25]

**CONCLUSIONS**

In conclusion, site specific cathodoluminescence spectroscopy and imaging in a scanning electron microscope are used to study the localized surface plasmon behavior of TOH Au NCs. The CL analyses combined with FDTD simulations have helped us to identify two dipole-active LSP modes. The higher wavelength (low energy) surface plasmon mode is governed by the dipolar mode of oscillations between the apexes of the particle, which is basically 3-fold degenerate in nature. The dipolar oscillation is also active for lower wavelength (higher energy) LSP mode but charge pattern oscillation takes place between base and apex points along the in-plane and out-of-plane direction. The proximity of the substrate with respect to the point of excitation also plays a role, especially on the degenerate apex-apex LSP mode of nanoparticle due to inherent anisotropy and complexity of the TOH Au particle. The remarkably high SERS response of the present TOH particles can be understood by invoking a general principle of energy conservation, for which the total electromagnetic energy per unit volume containing several nanoparticles should be same for any shape. Moreover, it is believed that hot spots between two close-by TOH particle surfaces (so called 'nanogaps') can be significantly stronger than single particle LSPRs indicating that MNPs aggregates may serve as reliable SERS substrate as compared to individual plasmonic nanoparticles for sensing and imaging applications.

## ASSOCIATED CONTENT

**Supporting Information**.

Comparison between standard dispersion response from Au and the FDTD fitted dispersion response, schematic of edge length and symmetric points of TOH particle, tilt angle HRSEM image of the TOH particles, reproducibility of the CL measurement, calculated extinction spectrum and additional SERS measurements are listed in the supporting information. This material is available free of charge via the Internet at http://pubs.acs.org.

**Corresponding Author**

*Tapas Kumar Chini.
* tapask.chini@saha.ac.in.

## ACKNOWLEDGMENT

We are grateful to the Department of Atomic Energy (DAE), Government of India, for the financial support. We are also grateful to the anonymous referees for the constructive criticisms of our paper. The authors thank Dr. Pabitra Das of Laboratoire de Physique des Solides, Université Paris Sud and Prof. Dulal Senapati of Saha Institute of Nuclear Physics for valuable suggestion and critical reading of the manuscript.

**REFERENCES.**

# Supporting Information

**Probing Localized Surface Plasmons of a Single Trisoctahedral Gold Nanocrystal and Demonstration of its Enhanced Raman Activity**

Achyut Maity[1], Arpan Maiti[1], Biswarup Satpati[1], Avinash Patsha[2], Sandip Dhara[2] and Tapas Kumar Chini*[1]

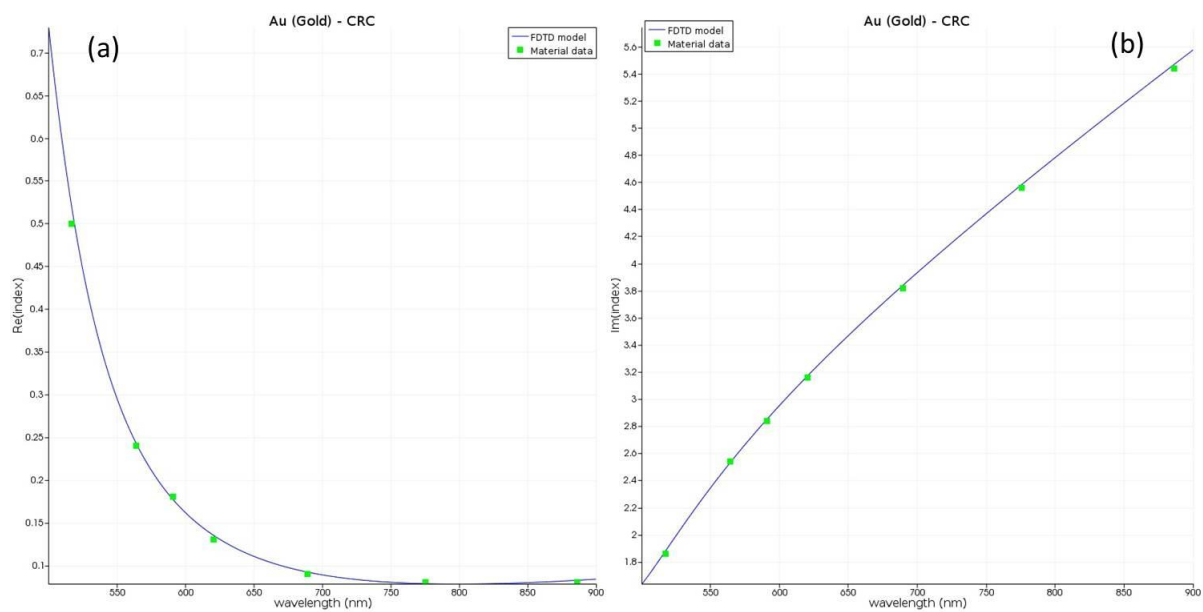

**Figure S1**: The material data obtained from the CRC Handbook of Chemistry and Physics[1] has been fitted with the (a) real and (b) imaginary part of refractive index for the modeled Au TOH particle using the multi-coefficient models (MCMs).

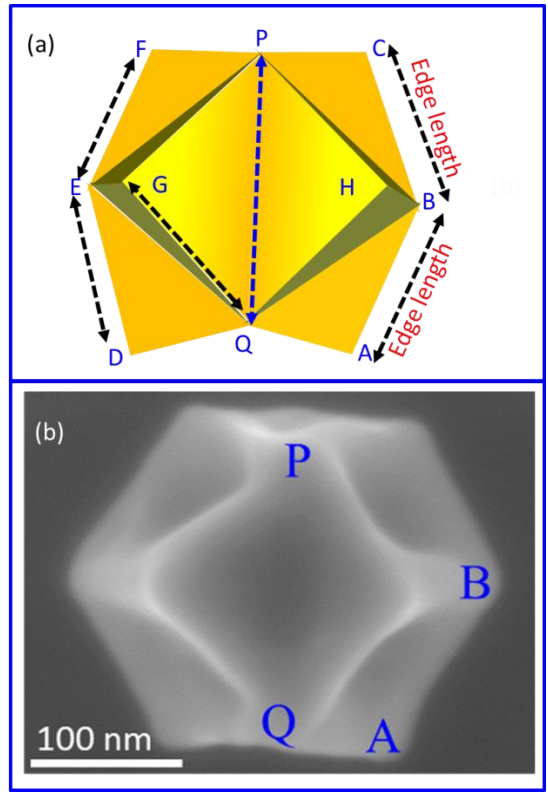

**Figure S2**: (a) Schematic of a typical concave TOH shaped particle. The apex points are marked as A, C, D, F, G and H (similar to the studied geometry) and the base points are marked as E and B. The edge length of a TOH particle is defined as the distance between base and apex points. Here the edge is either AB or BC. The HRSEM measurement (inset figure) shows that the edge length of the studied particle is ~ 125 nm. For simulation purpose we assumed this edge length to be 125 nm. The distance between the points P and Q is 190 nm (measured from HRSEM image, shown in (b)). This length is used as octahedron's edge length (also for the length of a side of triangular base of each pyramid) for simulation purpose. TOH particle has 2-fold rotational symmetry along the PQ plane. So, the point A, B and C are equivalent to the point D, E and F respectively at free-standing as well as at experimental configuration. When the TOH particle is rest on the Si substrate  Si substrate making an angle of 7.5o w.r.t the free-standing configuration, the points like A, D are nearer to the substrate than the points like B (or E), C (or F).

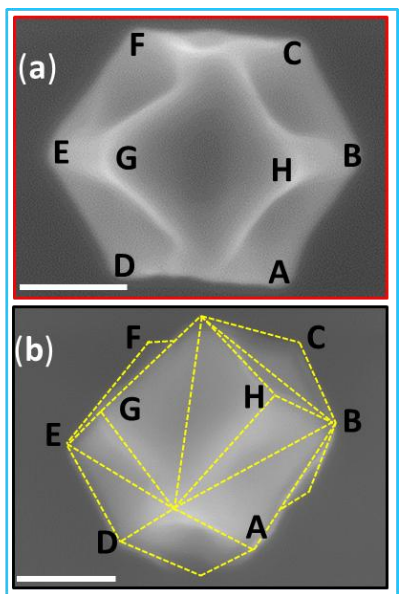

**Figure S3**: HRSEM image of the selected TOH particle at (a) 0° and (b) 45° tilt angle of the sample stage. Scale bar is 100 nm.

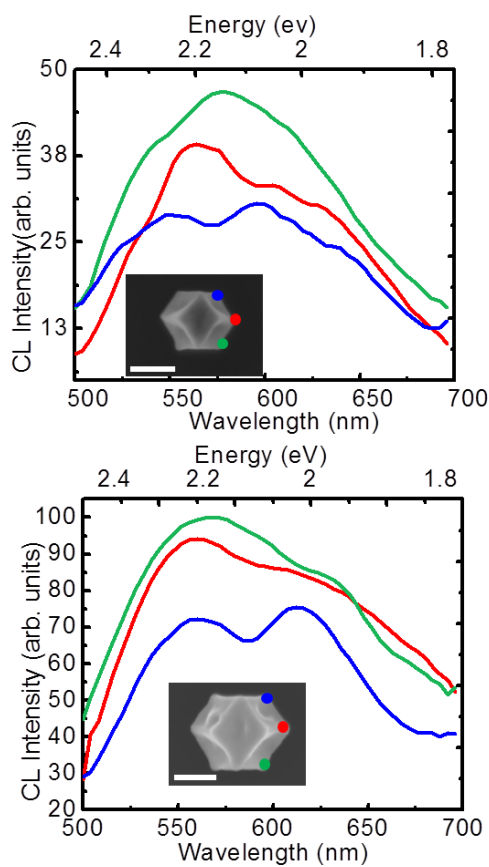

**Figure S4:** Experimentally acquired CL spectroscopic data from TOH particle of different sizes (in terms of edge length). The single and double peaked nature with sufficient broadening is very common in all the spectroscopic results and we have analyzed the best data in this present article. Scale bar is 100 nm.

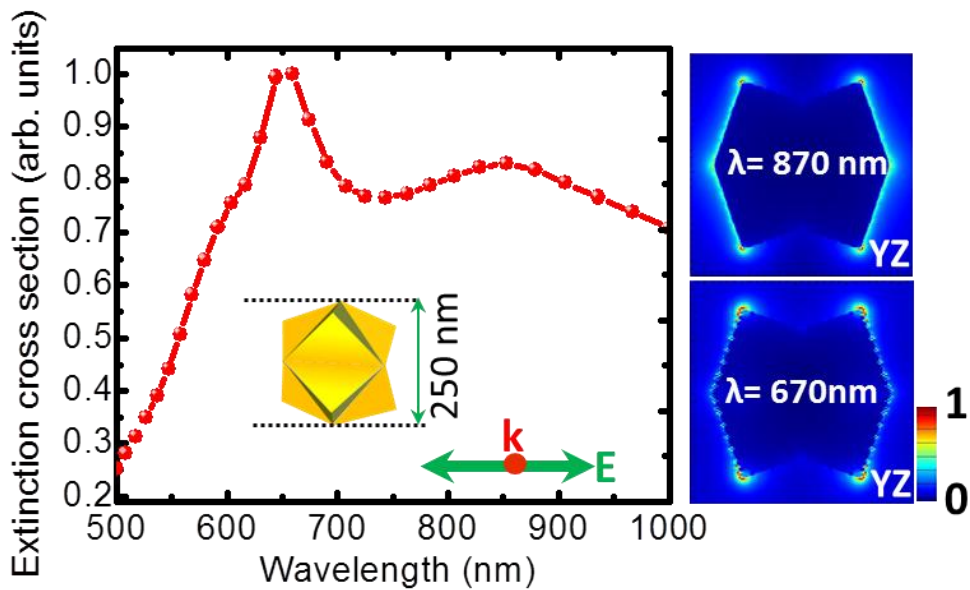

**Figure S5:** Calculated extinction spectrum and the corresponding field enhancement at 870 nm LSPR wavelength. The optical wave is propagating along X direction and the electric field vector is along the Y direction.

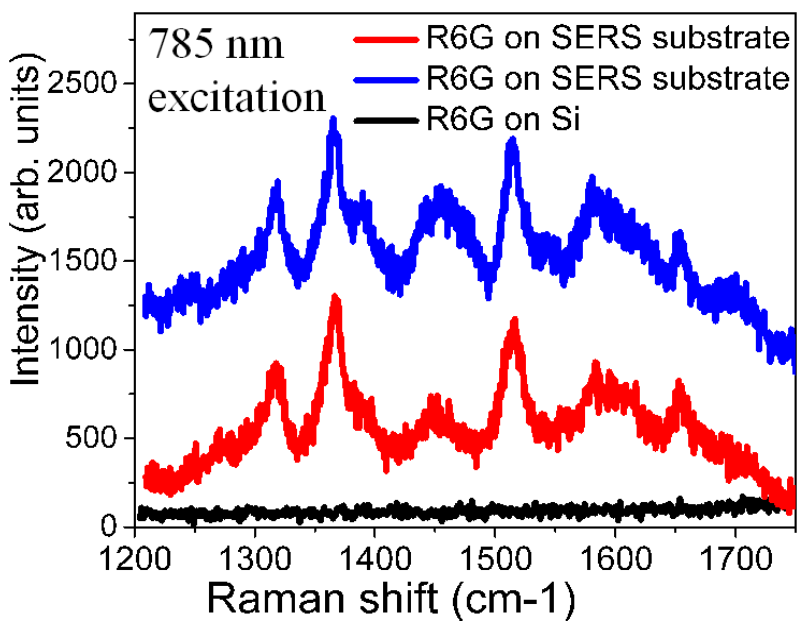

**Figure S6**: SERS spectra of R6G molecule (1μM) at different locations of the substrate, with excitation laser of wavelength 785 nm.

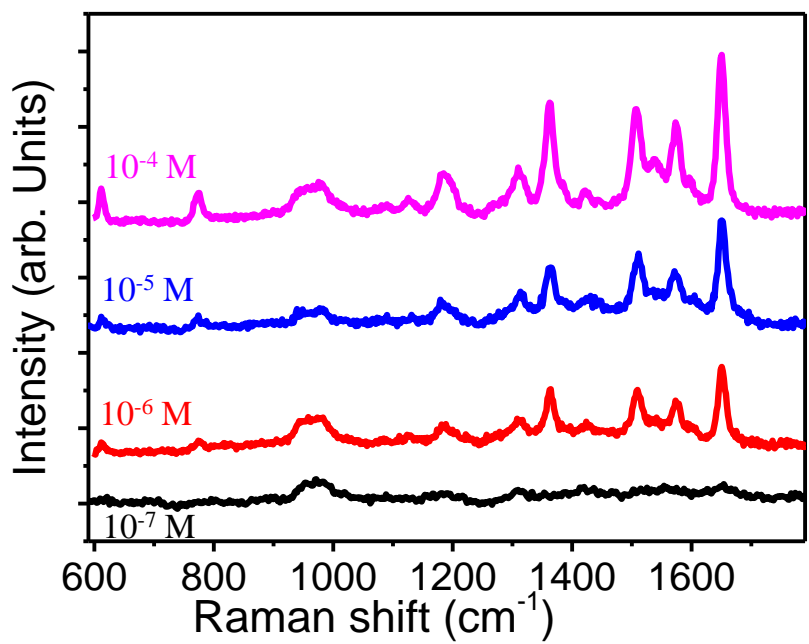

**Figure S7:** SERS spectra of R6G molecules in different aqueous solutions of concentrations ranging from $10^{-4}$ to $10^{-7}$ M, dispersed on SERS substrates.

| Marked positions / Information about resonant peaks | FWHM (in nm) of the lower wavelength (centered at around 548 nm) mode | FWHM (in nm) of the higher wavelength (centered at around 650 and 670 nm) mode |
|---|---|---|
| A | 80 | - |
| D | 75 | - |
| B | 60 | 170 (at 650 nm) |
| E | 60 | 183 (at 650 nm) |
| C | 58 | 190 (at 670 nm) |
| F | 53 | 193 (at 670 nm) |
| G | 60 | - |
| H | 73 | - |

**Table S1:** FWHM of experimentally acquired different plasmon modes.